\documentstyle[epsf,sprocl]{article}

\input{epsf}

\newcommand{\be}{\begin{equation}}
\newcommand{\ee}{\end{equation}}
\newcommand{\ba}{\begin{eqnarray}}
\newcommand{\ea}{\end{eqnarray}}

\newcommand{\vl}[1]{ \b{#1} }

\newcommand{\smslsh}[1]{#1 \! \! \! /}
\newcommand{\mydot}{\!\cdot\!}

\newcommand{\showfig}[2]{
\epsfxsize = #1
\centerline{\epsfbox{#2}}
}

\def\be{\begin{equation}}
\def\ee{\end{equation}}
\def\bea{\begin{eqnarray}}
\def\eea{\end{eqnarray}}
\begin{document}

\title{Non-Perturbative QED and QCD at Finite Temperature}

\author{Michael Strickland}

\address{Department of Physics, Ohio State University \\ OH, 43210, USA \\ Email:~~strickland.41@osu.edu}

%%%%%%%%%%%%%%%%%%%%%%%%%%%%%%%%%%%%%%%%%%%%%%%%%%%%%%%%%%%%%%
% You may repeat \author \address as often as necessary      %
%%%%%%%%%%%%%%%%%%%%%%%%%%%%%%%%%%%%%%%%%%%%%%%%%%%%%%%%%%%%%%

\maketitle\abstracts{ 
We present results of numerical solutions of Schwinger-Dyson equations for
the finite-temperature quark and electron propagators.  It is shown that both
strongly coupled QED and QCD undergo a chiral symmetry restoring
phase transition as the temperature is increased.  We go beyond the bare vertex or
``rainbow'' approximation by applying the finite-temperature 
Ward-Takahashi identity to constrain the non-perturbative vertex function.
}

\section{Introduction}

Within both strongly coupled quantum electrodynamics (QED) and quantum chromodynamics (QCD) there are analytic and numerical results which indicate that there is dynamical chiral symmetry breaking in these theories.\cite{miransky1,karsch}
At finite temperature it is widely believed that QCD undergoes a phase transition from the low-temperature phase with confined quarks and gluons and chiral symmetry breaking, to a high-temperature phase in which quarks and gluons are deconfined and chiral symmetry is restored.  Although the deconfinement and chiral phase transitions are two separate transitions, lattice studies find that the two transitions occur at approximately the same temperature.  Like QCD, strongly coupled QED is expected to undergo a finite-temperature phase transition which restores chiral symmetry.  The main tool for the theoretical investigation of non-perturbative finite-temperature QCD has been lattice gauge simulations \cite{karsch,rothe}.  However, the results from these studies are limited to small lattice sizes due to the numerical complexity of lattice simulations. In addition, extraction of detailed information (eg, bound-state wavefunctions) is complicated by the statistical nature of the data.  Another approach is to solve truncations of the finite-temperature Schwinger-Dyson (FTSD) hierarchy to obtain information about the phase structure of QED, QCD, and other field theories.$^{4-8}$
%\cite{craig,odintsov,cabo,alkofer,aitchison}

In this work, we study solutions of the FTSD equation for the electron and quark two-point functions in the imaginary-time formalism.  We extend previous treatments by using a vertex function which satisfies the finite-temperature Ward-Takahashi identity. Our results indicate that both strongly coupled QED and QCD undergo a chiral symmetry restoring phase transition.  The organization of the paper is as follows:  In Section II we present the finite-temperature formalism.  In Section III results of the numerical solution of the FTSD equation are presented and discussed.  Conclusions are drawn in Section IV.
  
\section{Finite Temperature SD Equations}

In the imaginary time formalism the finite-temperature Schwinger-Dyson equation is

\ba
\Sigma(p,u) &=& S^{-1}(p,u) - S_o^{-1}(p,u)
                        \nonumber \\
                    &=& \bar{g}^2
                        \int_k \gamma_{\mu} S(k,u) D_{\mu \nu}(p-k,u) \Lambda_\nu(k,p,u;p-k) \, ,
\label{ftsd}
\ea
where $\int_k = 1/\beta \sum_m \int d^d k / (2 \pi)^d$, $\bar{g}^2 \equiv g^2 (N_c^2-1)/(2 N_c)$, and $u$ is the heat bath four-velocity.  The sum is over odd Matsubara frequencies.\footnote{The QED FTSD equation is obtained when $\bar{g} \rightarrow e$.}

\subsection{Finite temperature quark and gluon propagators}

When written in covariant form, the most general finite-temperature quark
propagator is
\be
S^{-1}(p,u) = i A(p^2,p \mydot u) \smslsh{p} + B(p^2,p \mydot u) +
                i C(p^2,p \mydot u) \smslsh{u} \, .
\label{ftqp}
\ee
A tensor term proportional
to $\sigma_{\mu\nu}$ is ruled out by ${\mathcal P}{\mathcal T}$ invariance.\cite{furnstahl}

The finite-temperature gauge boson propagator in a covariant gauge,
assuming $q_\mu \Pi^{\mu \nu} = 0$, has the general form

\be
D^{\mu\nu}(q,u) = {\Delta_{3T}(q^2,q\cdot u) \over q^2 } P^{\mu\nu}_{3T} +
                         {\Delta_{3L}(q^2,q\cdot u) \over q^2 } P^{\mu\nu}_{3L} +
                         {\alpha \over q^2} P^{\mu\nu}_{4L} \, ,
\label{ftgp}
\ee
where $\alpha$ is the gauge fixing parameter, $P^{\mu\nu}_{4L}$ is the four dimensional longitudinal projector, 
and $P^{\mu\nu}_{3T}$ and $P^{\mu\nu}_{3L}$ are three dimensionally transverse and longitudinal
projectors, respectively.\cite{kapusta}

\subsection{Finite Temperature Rainbow Approximation}

If we replace $\Lambda_\nu$ by $\gamma_\nu$, the bare vertex, in (\ref{ftsd}) we get 

\be
\Sigma(p,p \cdot u) = \bar{g}^2 
                        \int_k \gamma_{\mu} S(k,k \cdot u)
                        D_{\mu \nu}(p-k,(p-k)\cdot u) \gamma_\nu \, .
\label{ftsdr}
\ee
Inserting the general forms of the finite-temperature quark and gluon propagators
and taking traces of (\ref{ftsdr}) with $- i \smslsh{p}$, $- i \smslsh{u}$, and $I$ 
gives three coupled integral equations
\ba
\Sigma_A &=& \left( {\mathcal A} - (p \cdot u) \, {\mathcal C} \right) / \vl{p}^2 \nonumber \\
\Sigma_B &=& \mathcal{B} \nonumber \\
\Sigma_C &=& \left( p^2 {\mathcal C} - (p \cdot u) \, {\mathcal A} \right) /  \vl{p}^2 \, ,
\label{ftreqs2}
\ea
where $\Sigma_A \equiv A - Z_2$,  $\Sigma_C \equiv B - M_o$, and $\Sigma_C \equiv C$, and
\ba
{\mathcal A} &\equiv& {\rm Tr[} \, - i \smslsh{p} \, \Sigma(p,p \cdot u) \, {\rm ]}
        \nonumber \\
{\mathcal B} &\equiv& {\rm Tr[} \, {\bf 1} \, \Sigma(p,p \cdot u) \, {\rm ]}
        \nonumber \\
{\mathcal C} &\equiv& {\rm Tr[} \, - i \smslsh{u} \, \Sigma(p,p \cdot u) \, {\rm ]}
        \, .
\label{ftreqs3}
\ea
When the traces in (\ref{ftreqs3}) are performed we get
\ba
{\mathcal A} &=& { \bar{g}^2 T \over (2 \pi)^2 } \sum_{m=-\infty}^{\infty}
                \int_0^\infty \vl{k}^2 d \vl{k} \left(
                { K_A(\vl{p},\omega_n,\vl{k},\omega_m;A,C) \over
                  A^2 (\vl{k}^2 + \omega_m^2) + 2 A C \omega_m + B^2 + C^2 }
                  \right) \nonumber \\
{\mathcal B} &=& { \bar{g}^2 T \over (2 \pi)^2 } \sum_{m=-\infty}^{\infty}
                \int_0^\infty \vl{k}^2 d \vl{k} \left(
                { K_B(\vl{p},\omega_n,\vl{k},\omega_m;B) \over
                  A^2 (\vl{k}^2 + \omega_m^2) + 2 A C \omega_m + B^2 + C^2 }
                  \right) \nonumber \\
{\mathcal C} &=& { \bar{g}^2 T \over (2 \pi)^2 } \sum_{m=-\infty}^{\infty}
                \int_0^\infty \vl{k}^2 d \vl{k} \left(
                { K_C(\vl{p},\omega_n,\vl{k},\omega_m;A,C) \over
                  A^2 (\vl{k}^2 + \omega_m^2) + 2 A C \omega_m + B^2 + C^2  }
                  \right) \, ,
\label{ftreqs4}
\ea
where $K_A, K_B,$ and $K_C$ are the integration kernels which depend on the form
of $\Delta_{3L}$ and $\Delta_{3T}$ and $\vl{k}^2 \equiv k_1^2 +k_2^2 + k_3^2$.

\begin{figure}
\showfig{8cm}{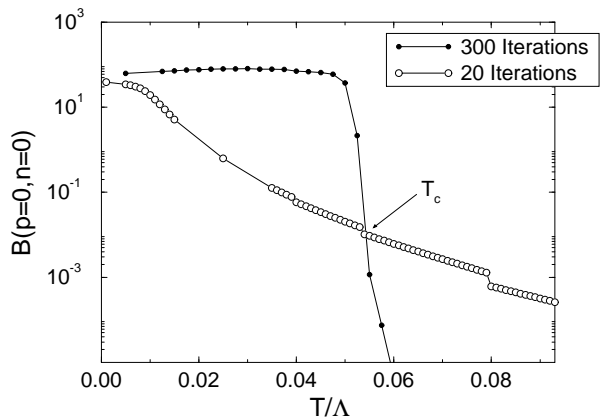}
\caption{$B(\vl{p}=0,n=0)$ as a function of $T/\Lambda$ showing the chiral symmetry restoration temperature $T_c (\bar{g}^2=20\,,\Lambda=1000)$.}
\label{crittemp}
\end{figure}

\subsection{Finite Temperature Quenched Rainbow Approximation}

When the gluon or photon propagator is taken to be bare ($\Delta_{3L} =
\Delta_{3T} = 1 = \Delta_{4T}$) the kernels in equations (\ref{ftreqs4})
can be evaluated.\cite{mydiss}
The resulting equation can be solved
iteratively by evaluating the integrals using Gaussian quadratures and
performing the Matsubara sums numerically.  The temperature dependence of the
scalar coefficient function is shown in Fig.\ref{crittemp}.  We
see that the mass is relatively unchanged until $T$ just below the critical
temperature for chiral symmetry restoration, at which point it undergoes a very
sharp transition into the symmetric phase. This suggests that the chiral phase 
transition is first order in quenched QED/QCD.  

Note here that the discontinuities in $B$ as a function of $T$ seen in Fig. \ref{crittemp} are a result of the finiteness of the UV cutoff, $\Lambda$.  Since $\omega_{n_{\rm max}} \sim \Lambda$, as the temperature increases fewer modes are included in the Matsubara sum.  If the UV cutoff, $\Lambda$, were significantly larger, then these discontinuities would be much smaller, and as $\Lambda \rightarrow \infty$ these discontinuities would be removed completely.  We are unable to take $\Lambda$ very large due to the computer time needed to compute the Matsubara sums.

\begin{figure}
\showfig{7cm}{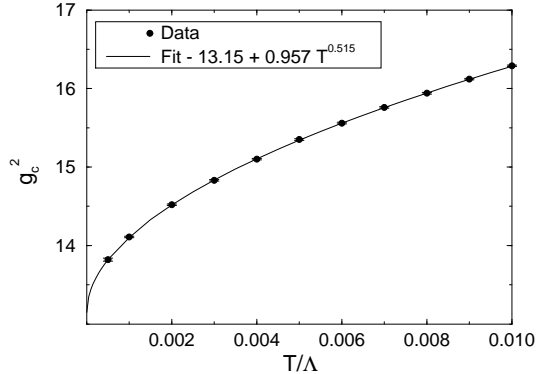}
\caption{Critical coupling $g_c^2$ as a function of temperature along with
fit of the form $g_c(T) = A + B T^C$ including the effects of the field
renormalizations ($\Lambda=$ 1000).}
\label{critcoup2}
\end{figure}

Using this method, we have calculated the temperature dependence of the
critical coupling for dynamical mass generation.  
The result of this calculation is given in Fig. \ref{critcoup2} showing that the critical coupling increases with temperature.  Therefore, for fixed coupling, $g_o$, above the critical coupling there is a finite-temperature chiral phase transition when $g_c(T) > g_o$.

\subsection{The Finite Temperature Vertex Function}

The requirement that physical quantities be independent of the choice of gauge leads to
non-perturbative relations between the fermion propagator and the vertex function, $\Lambda_\nu$.
Within QCD this relation is called the Slavnov-Taylor (ST) identity\cite{itzykson,hawes}
\ba
(p-q)_\nu \Lambda^a_\nu(p,q;p-q)[1 + b(q^2)] &=& \hspace{4cm} \, \nonumber \\
        &\,& \hspace{-5cm} [g T^a - B^a(p,q;p-q) ] S^{-1}(p) -
                S^{-1}(q)  [g T^a - B^a(p,q;p-q) ] \, ,
\ea
where $a = 1 \ldots N_c^2-1$ is the color index, $b(k^2)$ is the ghost
self-energy, and $B^a$ is the ghost-quark scattering kernel.  

For QED this identity reduces to the abelian version of the ST identities, the
Ward-Takahashi (WT) identity\footnote{For axial gauges, there are no ghost fields and therefore, within these gauges, the ST and WT identities are equivalent.}
\be
i (p-q)_\nu \Lambda_\nu(p,q;p-q) = S^{-1}(p) - S^{-1}(q) \, .
\label{wtid}
\ee

The most general
form for the finite-temperature fermion-gauge boson vertex function is
\ba
\Lambda_\nu(p,k;p-k)
        &=&  \Lambda_\nu^L(p,k;p-k) + \Lambda_\nu^T(p,k;p-k) \nonumber \\
        &=& (p+k)_\nu \left[ a(p,k,u)(\smslsh{p} + \smslsh{k})
                + b(p,k,u) + c(p,k,u) \smslsh{u} \right] \nonumber \\
        &\,&\hspace{.1cm} + d(p,k,u)\gamma_\nu
		+ u_\nu \left[ e(p,k,u) (\smslsh{p} +
                \smslsh{k}) + f(p,k,u) + g(p,k,u) \smslsh{u} \right]
                \nonumber \\
        &\,&\hspace{.1cm} + \Lambda_\nu^T(p,k;p-k) \, ,
\label{genftvertex}
\ea
Inserting the finite-temperature fermion propagator (\ref{ftqp}) and vertex function (\ref{genftvertex}) into (\ref{wtid}) gives four equations.
Unfortunately we have seven unknowns so this is not enough to solve for all of the coefficient functions that appear in the finite-temperature vertex function. We can, however, find
a one-parameter solution which satisfies the finite-temperature Ward-Takahashi (FTWT) identity:
\ba
a(p,k,u) &=& \bar{a} \; { A(p^2,p \mydot u) - A(k^2,k \mydot u) \over
                        2 (p^2 - k^2) } \nonumber \\
b(p,k,u) &=& \bar{a} \; { B(p^2,p \mydot u) - B(k^2,k \mydot u) \over
                        i (p^2 - k^2) } \nonumber \\
c(p,k,u) &=& \bar{a} \; { C(p^2,p \mydot u) - C(k^2,k \mydot u) \over
                        (p^2 - k^2) } \nonumber \\
d(p,k,u) &=& {1 \over 2} \left( A(p^2,p \mydot u) + A(k^2,k \mydot u) \right) \nonumber \\
e(p,k,u) &=& (1-\bar{a}) { A(p^2,p \mydot u) - A(k^2,k \mydot u) \over
                        2 u\mydot(p - k) } \nonumber \\
f(p,k,u) &=& (1-\bar{a}) { B(p^2,p \mydot u) - B(k^2,k \mydot u) \over
                        i u\mydot(p - k) } \nonumber \\
g(p,k,u) &=& (1-\bar{a}) { C(p^2,p \mydot u) - C(k^2,k \mydot u) \over
                        u\mydot(p - k) } \, .
\ea
The solutions for $e$, $f$, and $g$ are divergent for $n=m, p \neq k$.  Choosing $\bar{a}=1$ so that
all of the coefficient functions are finite we obtain the following solution to the FTWT identity

\ba
a(p,k,u) &=& { A(p^2,p \mydot u) - A(k^2,k \mydot u) \over
                        2 (p^2 - k^2) } \nonumber \\
b(p,k,u) &=& { B(p^2,p \mydot u) - B(k^2,k \mydot u) \over
                        i (p^2 - k^2) } \nonumber \\
c(p,k,u) &=& { C(p^2,p \mydot u) - C(k^2,k \mydot u) \over
                        (p^2 - k^2) } \nonumber \\
d(p,k,u) &=& {1 \over 2} \left( A(p^2,p \mydot u) + A(k^2,k \mydot u) \right) 
			\nonumber \\
e(p,k,u) &=& f(p,k,u) = g(p,k,u) = 0 \, .
\label{ftvf}
\ea

\begin{figure}
\showfig{8cm}{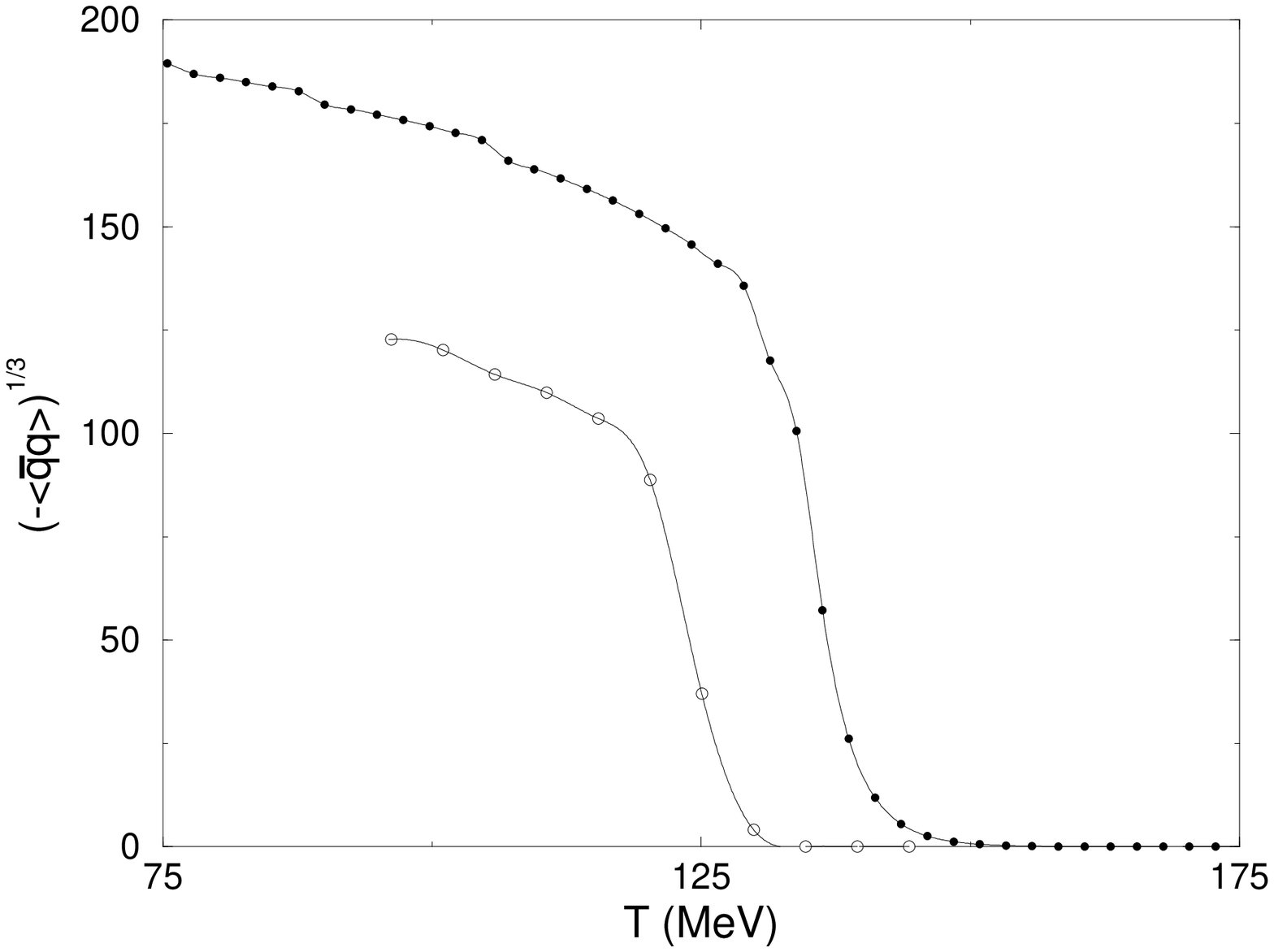}
\caption{Quark condensate $(-< \! \bar{q} q \! >)^{{1 \over 3}}$ as a function of temperature with the bare vertex ($\circ$) and the improved vertex ($\bullet$) Eq. (\ref{ftvf}). ( ($m_t^2=.69\,{\rm GeV}$, $\bar{g}^2=17.546$, $M_o=1\,{\rm MeV}$, $\Lambda=$ 10\,{\rm GeV})}
\label{condensate}
\end{figure}

\subsection{Temperature Dependence of the Quark Condensate}

To exactly solve the Schwinger-Dyson equations for the quark propagator we must
also solve the coupled equations for the gluon propagator and
all of the higher $n$-point functions.  Short of this we can
propose analytic forms for the gluon propagator and study the resulting quark
propagators.  At zero-temperature there has been much work along these lines.$^{14-21}$
%\cite{roberts1,burden,nemirovsky,maris,maris1,mckay,habel1,habel2}

Since perturbative renormalization group studies
are not reliable below $p^2 \sim$ 1 ${\rm GeV}^2$ the running of the coupling
constant at low energy must be determined in a different manner.  Results from lattice and
analytic studies suggest that the zero-temperature gluon propagator is significantly enhanced for
small spacelike-$k^2$.$^{22-24}$
%\cite{smekal,bender,atkinson} 
Traditional potential
analysis suggests that the gluon propagator behaves like $1/q^4$ for small $q$ and
studies have been performed with this assumption or regularizations of this form.$^{24-27}$
%\cite{atkinson,jain,munczek2,munczek3} 
When modeling confinement we will instead use a parameterization which is a regularized version of $1/q^4
$ plus a perturbative piece
\be
{ \bar{g}^2 \Delta_{3L}(q,\omega_n)  \over \vl{q}^2 + \omega_n^2 } = { \bar{g}^2 \Delta_{3T}(q,\omega_n)  \over 
\vl{q}^2 + \omega_n^2 } = m_t^2 \, {(2 \pi)^3 \over T} \,
\delta^3(q)
        \delta_{n0} + {4 \pi^2 d \over \vl{q}^2 + \omega_n^2} \, ,
\label{ftmgp}
\ee
where $d = 12/(33 - 2 N_f)$.  This parameterization contains two pieces: a
delta function which models long-range effects like confinement through an
integrable infrared singularity, and a perturbative piece which ensures that the
large-$q^2$ limit is correct up to logarithmic corrections.\cite{craig}  
The parameter $m_t$ can be interpreted as the energy scale which separates
the perturbative and non-perturbative regimes.  In a zero-temperature 
calculation using a similar parameterization, Roberts
and Frank have fixed $m_t$ by fitting to a set of pion observables (eg, $f_\pi$, $N_\pi$
, scattering lengths) and find $m_t = 0.69\,{\rm GeV}$.\cite{roberts1} 

Once the gluon two-point function is specified, the FTSD equation with the 
vertex function (\ref{genftvertex},\ref{ftvf}) gives the temperature dependence of $A$, $B$, and $C$.
The quark condensate can be written in terms of these functions
\be
< \! \bar{q} q \! > = -12 \int_k \left( {B(k^2,\omega_m) \over A^2 p^2 + 2 A C \omega_m + B^2 + C^2} - { M_o \over p^2 + M_o^2 } \right) \, .
\ee

In Fig.\ref{condensate} we plot the temperature dependence of the quark condensate within the rainbow approximation and with the improved vertex (\ref{genftvertex},\ref{ftvf}).  The magnitude of the condensate and the critical temperature depend sensitively on the vertex used, demonstrating that proper treatment of the finite-temperature vertex function is necessary.  In particular, the rainbow approximation vertex function breaks gauge invariance and therefore the critical temperature will depend on the choice of gauge.  By using a vertex function (\ref{genftvertex},\ref{ftvf}) which satisfies the FTWT identity (\ref{wtid}) the gauge dependence of the critical temperature is reduced.

\section{Conclusions}

In this work we have shown that is it possible to solve truncated FTSD equation
beyond the bare vertex approximation.  The propagators obtained allowed us to study the nature of the chiral
phase transition which we find to be first-order when the bare gauge boson propagator is used.  
The coefficient functions 
which appear in the finite-temperature fermion propagator are determined non-perturbatively and can be
used to calculate other quantities like $m_\pi$ and $f_\pi$ in the chiral limit.\cite{craig}  When there is explicit chiral symmetry breaking, these propagators can be used in solutions of the 
finite temperature Bethe-Salpeter equation in order to obtain the temperature dependence of the masses and widths of the light to intermediate mesons.

\section*{Acknowledgments}

I would like to thank C. Roberts for helpful discussions.  This work was supported 
by the National Science Foundation under Grants No. PHY-9511923 and PHY-9258270.

\end{document}